\documentclass{iaus}
\usepackage{graphicx}

\title[Model atmospheres of transiting planets] 
{Spectrum and atmosphere models of irradiated transiting extrasolar giant planets}
\author[Ivan Hubeny \& Adam Burrows]   
{Ivan Hubeny$^1$
\and Adam Burrows$^2$}

\affiliation{$^1$Steward Observatory and Dept. of Astronomy, 
University of Arizona, \\ 933 N. Cherry Ave., 
Tucson, AZ 85721, USA \\ email: {\tt hubeny@as.arizona.edu} \\[\affilskip]
$^2$Dept. of Astrophysical Sciences, Princeton University, \\ Peyton Hall,
Princeton, NJ 0854, USA \\email: {\tt burrows@astro.princeton.edu}}

\pubyear{2008}
\volume{253}  
\pagerange{1-7}
\jname{Transiting Planets}
\editors{A.C. Editor, B.D. Editor \& C.E. Editor, eds.}
\begin{document}

\maketitle

\begin{abstract}
We show that a consistent fit to observed secondary eclipse data
for several strongly irradiated transiting planets demands 
a temperature inversion (stratosphere) at altitude.
Such a thermal inversion significantly influences the
planet/star contrast ratios at the secondary eclipse,
their wavelength dependences, and, importantly, the
day-night flux contrast during a planetary orbit.
The presence of the thermal inversion/stratosphere seems to 
roughly correlate with the stellar flux at the planet.
Such temperature inversions might caused by an upper-atmosphere 
absorber whose exact nature is still uncertain.

\end{abstract}

\firstsection 
\section{Introduction}
Theoretical modeling of atmospheres of extrasolar giant planets (EGP) 
is a young field, about one decade old, and as youngsters usually are,
it is very active, restless, and sometimes quite unpredictable.
It had recently undergone a transition from a purely care-free stage
(no observations were available) to a more difficult stage where there
already are some observations to be fit by the theory.

In this paper, we briefly describe our recent efforts in this
area. We will try to convince the reader that despite its young age
and many shortcomings, the theory is actually doing quite well.

\section{Modeling procedure}
The basic aim of a model atmosphere is to predict both the structure
(physical and chemical state) of an atmosphere as a function of depth,
as well as the theoretical emergent spectrum. The latter is the most
important result of a modeling exercise because by comparing the model
predictions to actual observations one might constrain the physical
conditions in planet atmospheres, and indirectly also of the whole planet.

The general problem of atmospheres of extrasolar giant planets is
an extremely complex one that should generally involve a 
3-dimensional radiation hydrodynamical treatment of atmospheric 
motions and day/night side
energy redistribution, with departures from chemical equilibrium,
and a self-consistent description of clouds (particle size
distribution, cloud position in the atmosphere), together
with a self-consistent treatment of the radiation field. Such a challenge is
still far beyond current computational capabilities, not withstanding the 
lack of, or inadequacies in, many ingredients needed for the modeling effort, 
such as complete and reliable atomic and molecular data and
optical constants for cloud particles, to name just a few.

Obviously, one has to resort to approximations. It is fair to say 
that the work of different groups differs mostly by what actual
approximation are adopted, and what simplifications are being made.
In the absence of exact calculations it is difficult to decide which
approximations are the most restrictive. Therefore, the current
and future development of this field will proceed by gradually relaxing more
and more approximations, and the observational tests, whenever
available, become the judge of whether one is on track.

The current standard model atmospheres of extrasolar planets are
based on the following assumptions: (i) the atmosphere is assumed
to be a plane-parallel, horizontally homogeneous layer (which reduces 
the problem to 1-D); 
(ii) the atmosphere is in hydrostatic equilibrium;
(iii) the atmosphere is in radiative equilibrium (or radiative+convective
equilibrium in convectively unstable layers);
(iv) local thermodynamic equilibrium (LTE) holds (therefore the opacity
and emissivity are known functions of only temperature and density, and
can, therefore, be pre-tabulated and not computed on the fly during 
model construction); and
(v) the atmosphere is in a modified chemical equilibrium (the local chemical
equilibrium which takes into account possible rainout and consequent
depletion of species).

Most groups that produce EGP model atmospheres adopt the above assumptions,
and use different modeling codes 
(e.g., \cite{SeSa98}, \cite{BAH01}).

Our group has originally used a code described
in \cite{Bur97}, with an additional algorithm to treat radiation
scattering in clouds --
\cite{SBP00}. We later switched to a more powerful procedure based on adapting
the general-purpose stellar atmosphere code TLUSTY 
(\cite{Hube88}; \cite{HL95}),
originally designed for non-LTE model atmospheres of hot stars, and
extended to other stellar types and accretion disks.
The specific variant for the atmospheres of EGP's and brown dwarfs is called
CoolTLUSTY, and has been described briefly in \cite{SBH03},
\cite{HBS03}, and \cite{BBH08}.
The code uses the above mentioned five basic assumptions. The chemical
equilibrium and departures from it due to the rainout of the species
is treated following \cite{BS99}, and the opacity
tables are computed as described in \cite{SB07}.

\section{Model bifurcation; existence of stratospheres}

Soon after the \cite{SBH03} grid of EGP model atmospheres was completed, 
the very close-in EGP, OGLE-TR-56b, (\cite[Konacki et al. (2003)]{Kon03})
was discovered, with a planet-star separation of a mere 0.0225 AU. We set out to extend 
our grid to higher irradiations, and found that the computed atmospheric
structure depends very sensitively on whether the TiO/VO opacity is
included in the opacity table or not -- see
\cite{HBS03}.
The basic conclusions of that paper are the following:
The models without TiO/VO always
exhibit a monotonic temperature/pressure ($T/P$) profile, with temperature
decreasing outward. When the TiO/VO opacity is included, the models remain
essentially unchanged for larger planet-star separations, simply because
TiO and VO are rained out everywhere in the atmosphere, save
in the deepest layers which have virtually no effect on emergent
spectra. For very close separations, the model with TiO/VO exhibits a significant
temperature increase toward the surface. In analogy with the solar-system planets,
the region of temperature inversion may be called a ``stratosphere." 
But the biggest surprise was that for a certain range of planetary
distances, there are actually two legitimate solutions when TiO/VO opacity is taken
into account -- one with a monotonic decrease of temperature toward surface and
one exhibiting a stratosphere! 

\begin{figure}
\begin{center}
 \centerline{
 \includegraphics[width=2.6in]{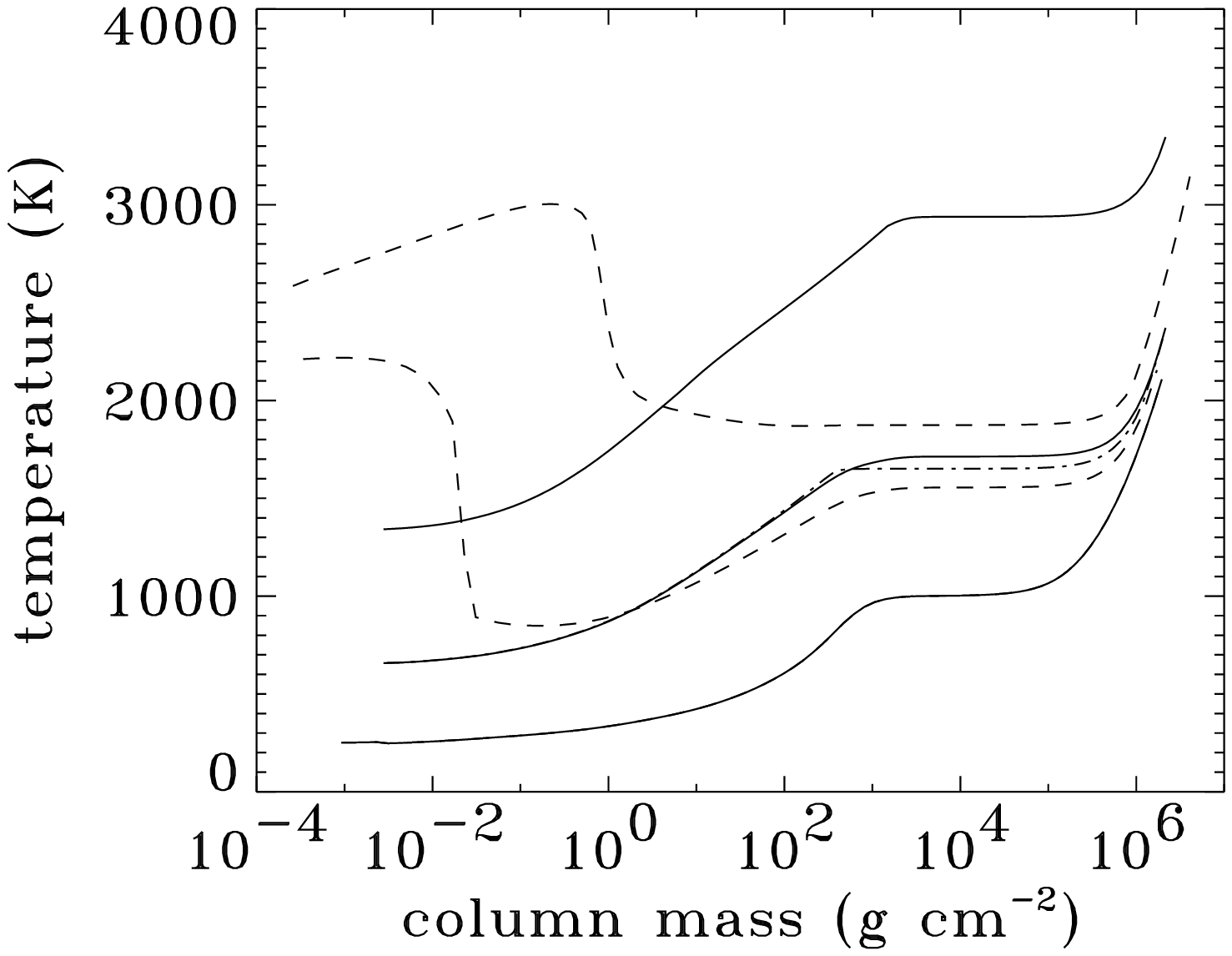} 
 \includegraphics[width=2.6in]{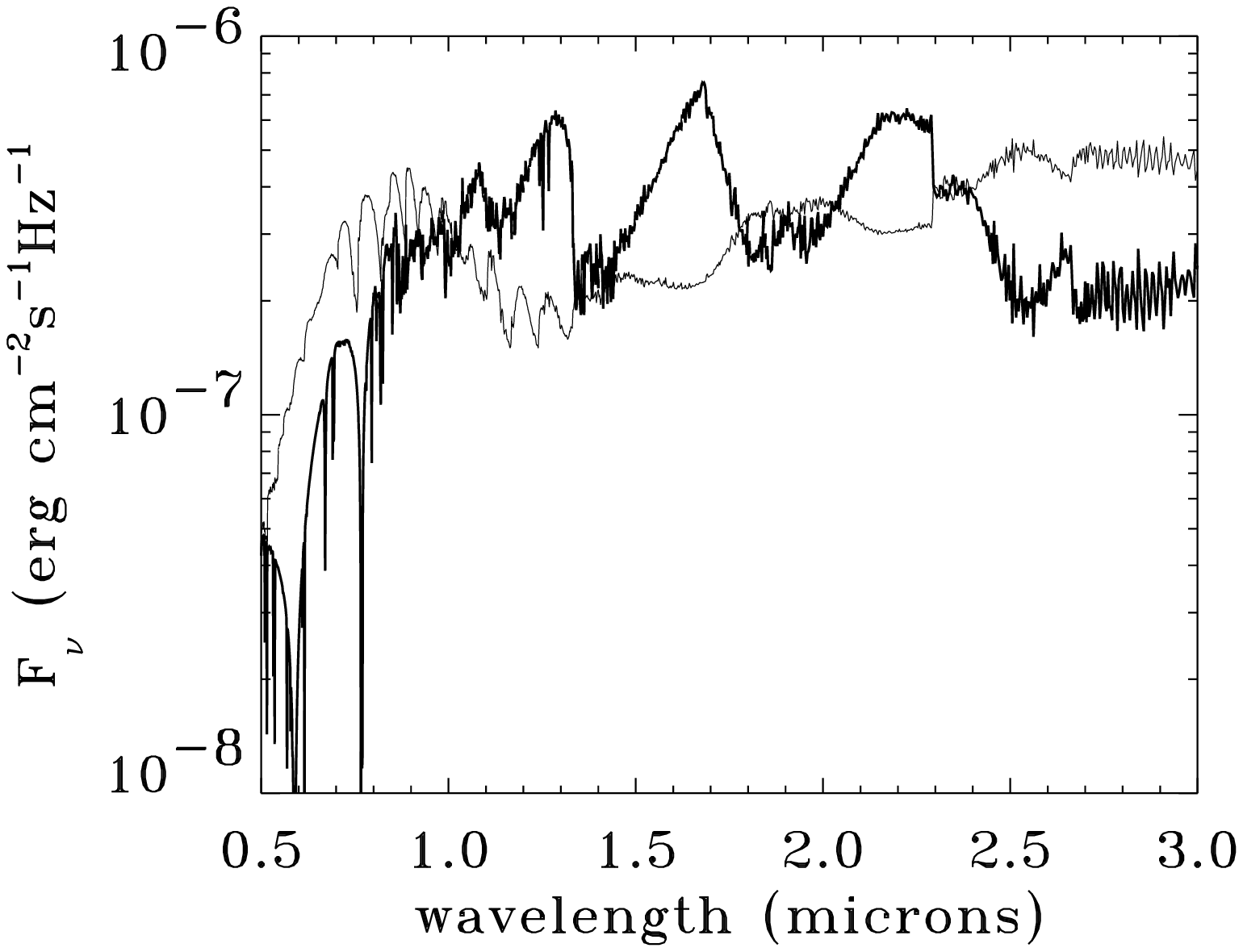} }
 \caption{Left panel:Temperature as a function of column mass.
Solid lines represent models computed for the opacity table without
TiO/VO, while the dashed and dot-dashed lines represent models 
with TiO/VO. The curves represent, from top to bottom,
three distances from the central star.
The top lines corresponds to 0.0225 AU, the middle lines to 0.08 AU,
and the bottom lines to 0.65 AU. 
The models for the lowest irradiation 
computed with and without TiO/VO are not distinguishable on the plot.
The middle dashed and the dot-dashed lines 
represent the two solutions for the same distance, and thus
illustrate the true bifurcation discussed in the text. 
Right panel: Emergent flux as a function of wavelength for
the high-irradiation (distance 0.0225 AU) model, computed
without TiO/VO (thick line) and with TiO/VO (thin line).
The models differ dramatically in the optical and in near-IR
region because of the dramatically different surface temperature
and the corresponding increase of the TiO/VO opacity. (Taken from Hubeny et al. 2003.)}
\label{fig1}
\end{center}
\end{figure}

This behavior is illustrated in Fig.~\ref{fig1}.
Full lines represent models computed for the opacity table without
TiO/VO, while the dashed and dot-dashed lines represent models 
with TiO/VO. The middle model exhibits the true bifurcation -- the
dashed and dot-dashed lines correspond to the two solutions for
the atmospheric structure with exactly the same input parameters
(effective temperature, surface gravity, elemental composition, and 
irradiation). The fact that the model without TiO/VO
is very close to the low-surface-temperature branch of the model 
with TiO/VO clearly
indicates that the model bifurcation, and the very existence of the
outward temperature increase, arises due to the TiO/VO opacity.

The physical explanation of this effect is straightforward -- see \cite{HBS03}.
TiO/VO provide a strong opacity in the optical wavelengths, where
the external irradiation (in the case of solar-type stars) has its
maximum. Therefore, the incoming stellar radiation is efficiently
absorbed in the upper atmosphere, thus leading to
its heating. The explanation of bifurcation is analogous. An atmosphere
either ``chooses'' a high surface temperature, in which case TiO/VO
is present, and the incoming stellar flux is efficiently absorbed
(so one indeed has, consistently, a high surface temperature). 
Or, the atmosphere ``choses'' a
low surface temperature, in which case TiO/VO does not exist near the
surface, and, therefore, there is no efficient mechanism to absorb
incoming stellar radiation in the upper atmosphere to heat it.

The difference in the temperature structure is reflected
in the emergent spectrum. In the right panel of Fig.~\ref{fig1}, we show the
emergent flux that coresponds to OGLE-TR-56, with a planet-star 
separation equal to 0.0225 AU. The thick line is a model 
without TiO/VO, and the thin line with TiO/VO. 
Similar results were later obtained by
\cite[Fortney et al. (2006)]{For06} and
\cite[Fortney et al. (2008)]{For08}.

However, there is a potentially serious problem with this picture.
Since the presence of TiO and VO is 
quite natural at high-$T$/high-$P$ points, one confronts a situation 
in which there are two regions rich in TiO/VO, separated
by a middle region that is TiO/VO free. In a gravitational field with a monotonic
pressure profile, this gap could act like a cold trap in which the TiO/VO that is
transported by molecular or eddy diffusion from the upper low-$P$ region into the
intermediate cooler region, would condense and settle out, thereby depleting the
upper low-$P$ TiO/VO-rich region. This would eventually leave no 
TiO/VO at altitude to provide
the significant absorption that could lead to a bifurcation. 

The situation is much more favorable for very close-in planets,
where the temperature is so high
that TiO/VO is present everywhere in the atmosphere. In this case,
one obtains a single solution, that with a stratosphere.
We note that the models presented in \cite{HBS03} considered no day/night
side heat redistribution. Consequently, these models generally
tend to overestimate the atmospheric temperature.

We conclude that
TiO/VO reliably operate only in a limited temperature range ($T > 1500$ K),
because otherwise the cold trap effect may deplete them.
One can speculate, though, that diffusion, turbulent mixing, or a wind may actually replenish TiO and VO,
but no explicit calculations or even rough estimates have yet been done.
Experience from the solar-system giant planets indicates that a number of unknown,
optically-active compounds may exist in the upper atmosphere.
With extreme irradiation, a complex non-equilibrium photochemistry will very likely
take place. Consequently,
there may be a number of as yet unknown opacity sources in the upper atmosphere that
may build a stratosphere, even in the absence of TiO/VO.
In view of these uncertainties, one is free (indeed, forced) to parameterize the extra
opacity; specifically its magnitude ($\kappa_e$), and its position in the 
physical and the wavelength space. Therefore,
we have adopted such a philosophy. Our approach is described in detail in
\cite{BBH08}.

\section{Day/night side heat redistribution}

In early studies (\cite{SBH03};  
\cite{BAH01};
\cite{BSH04};
\cite{BHS05};
\cite{BSH06}),
the degree of the day/night side heat redistribution was described 
through an empirical parameter, originally denoted as $f$.
In \cite{BSH06} we introduced an analogous parameter, $P_n$, as a fraction of incoming flux 
that is redistributed to the night side. The underlying assumption was
that the fraction $P_n$ of the incoming flux is
somehow removed before the incoming radiation reached the
upper boundary of the atmosphere, and deposited at
the lower boundary of the night-side atmosphere.

Although this treatment is useful for obtaining a rough estimate of the
effects of day/night side heat redistribution, it is unsatisfactory from
the physical point of view. There are two main problems:
(i) it is unphysical to assume that the incoming flux is removed at 
the top of the day-side atmosphere and injected in the bottom of the
night-side atmosphere, and (ii) the calculated entropies in the convection 
zones at the day and the night side may be generally different. Since 
convective transport is very efficient, the temperature gradient is
essentially exactly adiabatic. Consequently, the whole convective core
has to he isentropic, and, thus, cannot exhibit any difference between
the day and night sides.

To overcome these problems, we have recently adopted an improved,
albeit still parametric, approach -- \cite{BBH08}.
Out of the total incoming energy, the fraction $(1-P_n)$  is assumed to be radiated on the
day side, and the fraction  $P_n$  on the night side.
The irradiation energy is removed on the day side in a given depth range,
parameterized by limiting pressures $P_0$ and $P_1$.
This energy is deposited on the night side in depth range parameterized by
limiting pressures $P_0^\prime$ and $P_1^\prime$, which are generally 
different from $P_0$ and $P_1$, but in actual calculations we usually 
take them to coincide with $P_0$ and $P_1$. The modeling proceeds as follows:
at the day side, the input parameters are
$\log g$ ($g$ being the surface gravity); $T_{\rm int}$ 
(which corresponds to $T_{\rm eff}$ in the case of stellar atmospheres); 
$P_n$; $P_0$; and $P_1$. Other parameters are the distance and the radius
of the parent star, and the elemental composition (which is assumed to be the same
on the night side).
The computed model gives the structure (essentially, the $T/P$ profile and
the self-consistent radiation intensity), plus the entropy in the convection zone.
At the night side, the
input parameters are: $\log g$ (which must be the same as at the day side); 
$P_n$; limiting pressures $P_0^\prime$ and $P_1^\prime$; 
and a trial $T_{\rm int}^{\rm night}$. The model again yields a $T/P$ profile,
and the entropy at the convection zone, which is generally different from the
entropy at the day side.
The intrinsic temperature $T_{\rm int}^{\rm night}$ is adjusted by trial 
and error so that both entropies (day and night) match. We note that
with strong irradiation, the convection zone may be very deep
(at $\tau_{\rm ross}$ of the order of $10^4$ to $10^6$), and, therefore,
the entropy matching is important for evolutionary models, and not so much 
for predicting emergent spectra.

\section{Theoretical interpretation of the secondary eclipse spectra}

For close-in EGPs the planet-star contrast ratios in the mid-infrared
are relatively large, often exceeding $10^{-3}$, and such contrasts
are well within reach of the infrared space telescope {\it Spitzer}.
Using its IRAC and MIPS cameras, and the IRS spectrometer, one can now
measure the summed light of the planet and the star in and out of the
secondary eclipse, and, from the difference, determine the planet's
spectrum at superior conjunction 
(see several reviews in these Proceedings).
This has led to a breakthrough
in the study of extrasolar planets, for it provides a means to probe
the physics, chemistry, and even meteorology of their atmospheres.

Before this conference, secondary-eclipse fluxes in the IRAC and MIPS channels
had been measured for five transiting planets -- HD189733b, TrES-1,
HD 209458b, HD 149026b, and GJ 436b. Our group has performed a detailed
analysis of four of them (\cite{BBH08}).
So far, we have not analyzed GJ 436b.

\begin{figure}

\centerline{
\includegraphics[width=5.cm,angle=-90,clip=]{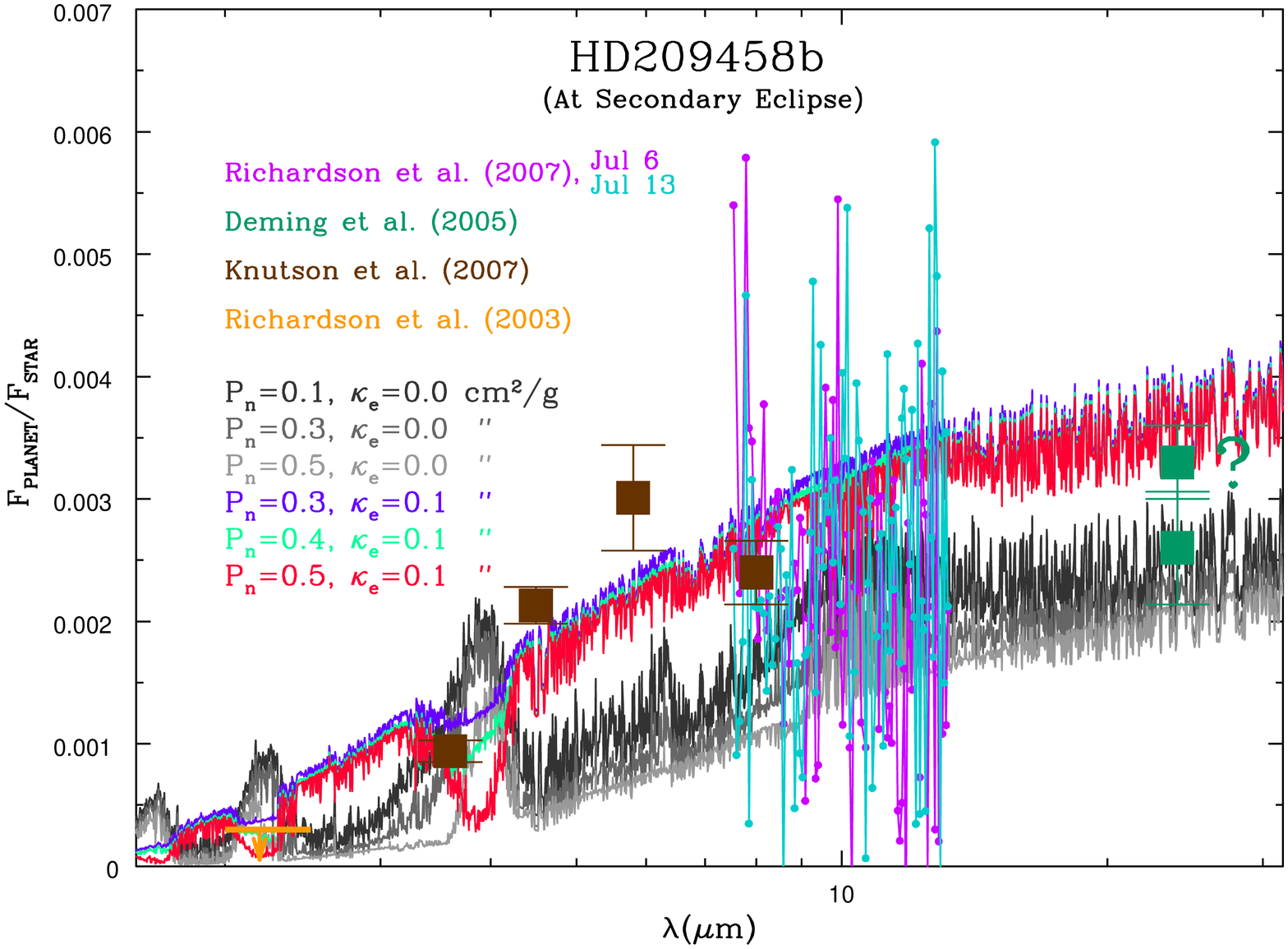}
\includegraphics[width=5.cm,angle=-90,clip=]{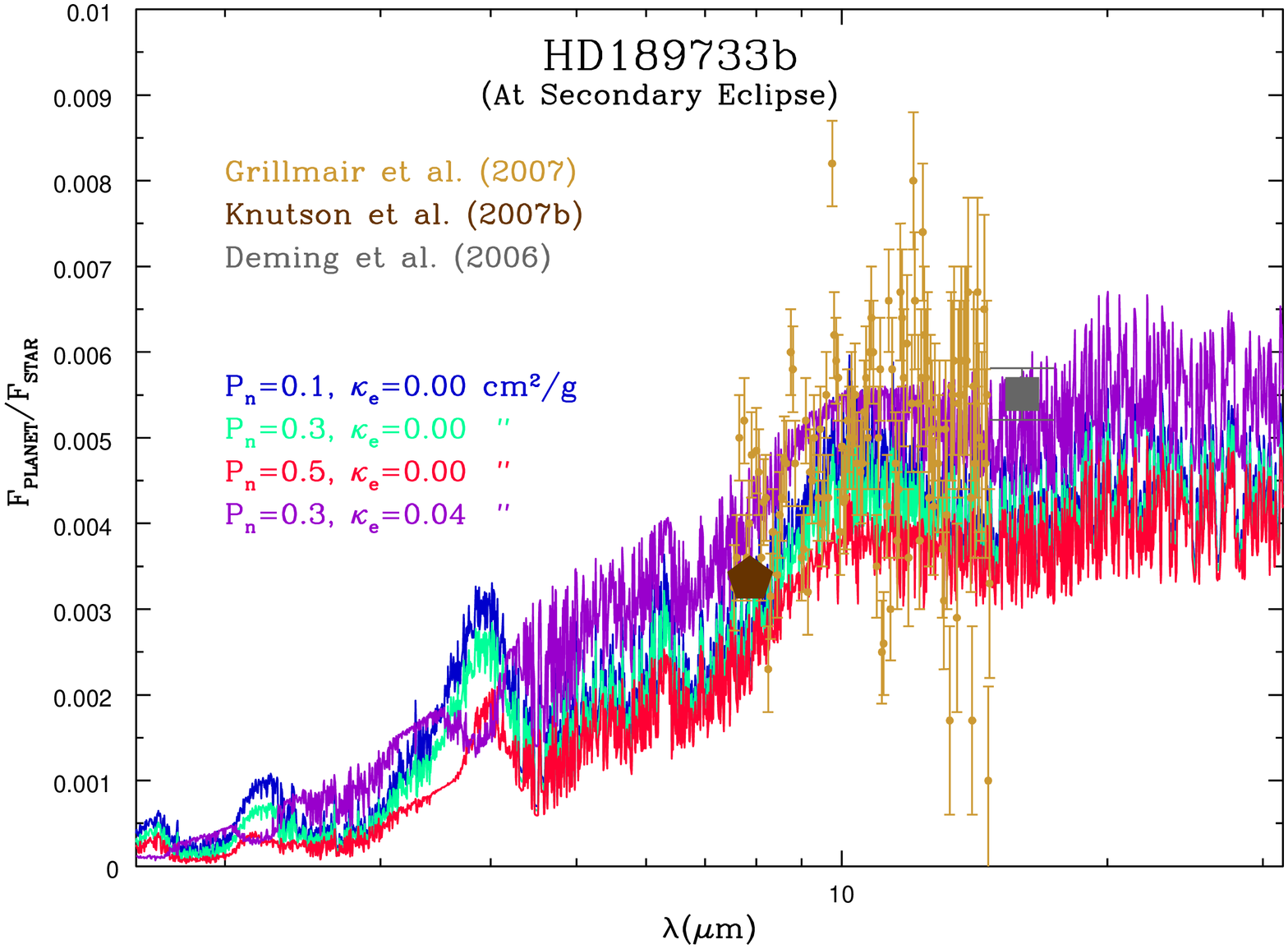}}
\centerline{
\includegraphics[width=5.cm,angle=-90,clip=]{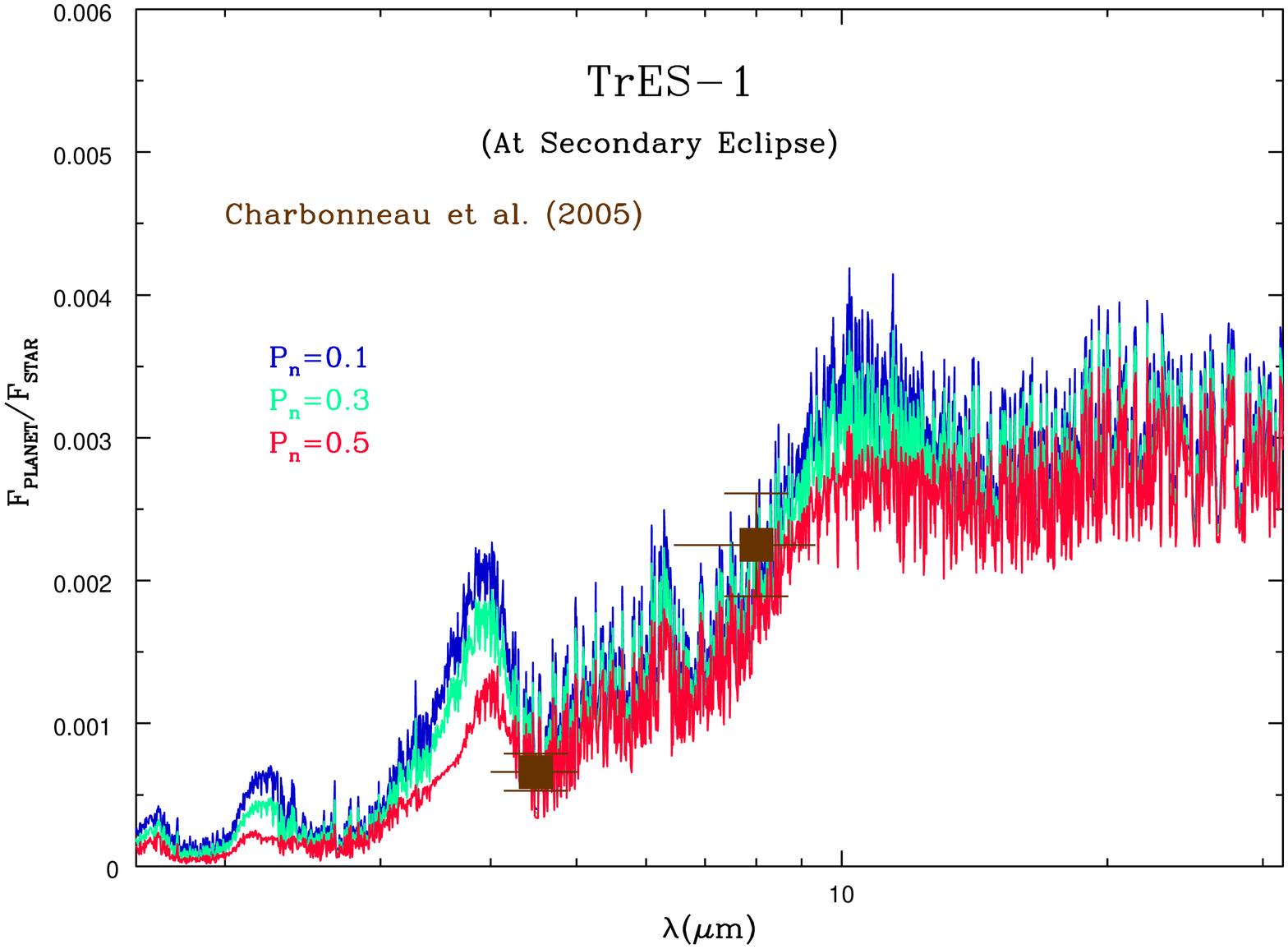}
\includegraphics[width=5.cm,angle=-90,clip=]{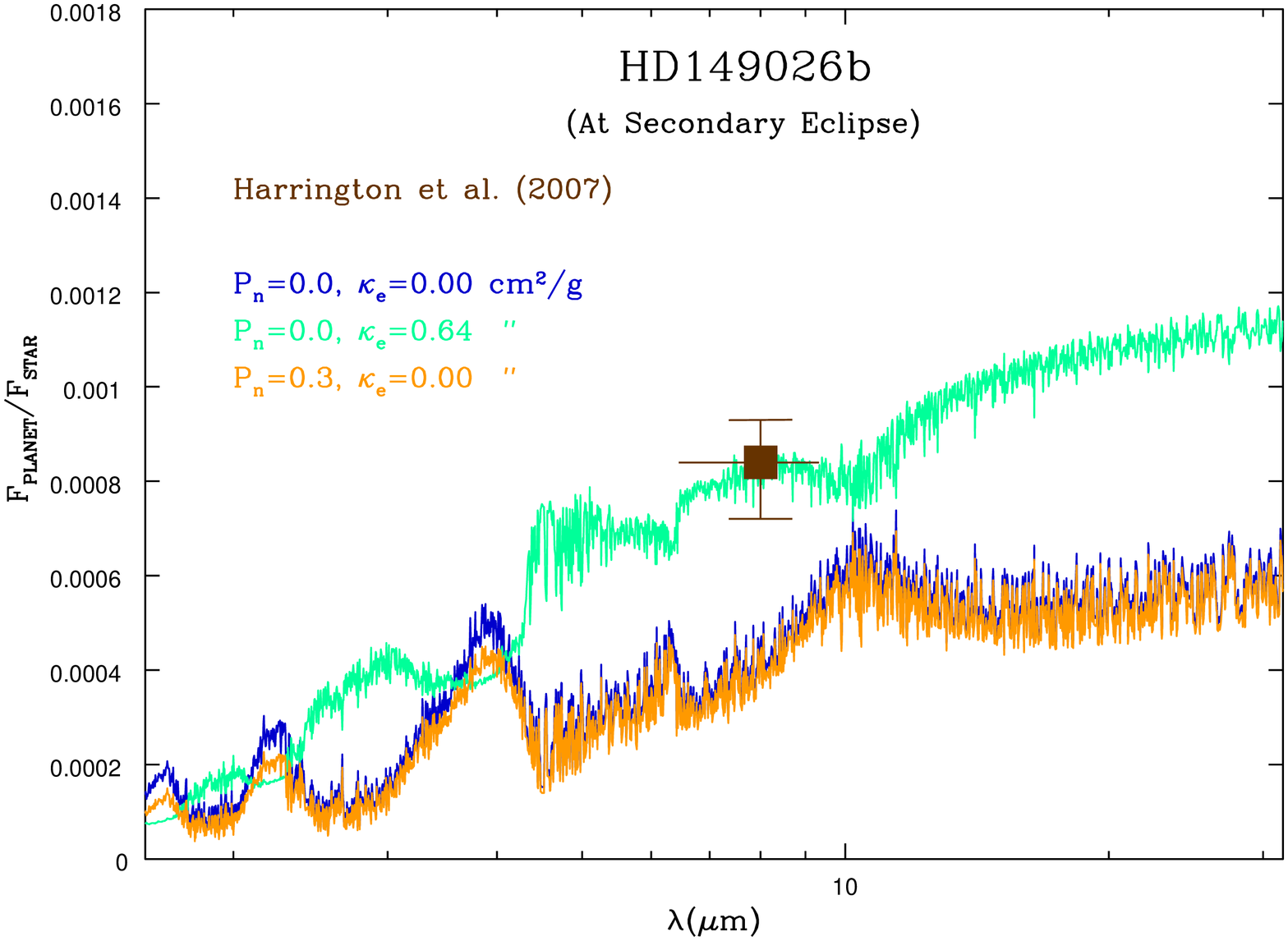}}
\caption{
The planet/star flux ratios versus for various models of four transiting
EGPs measured by {\it Spitzer} at secondary eclipse.
Notice the different scales employed in each panel.  Models
for different values of P$_n$ and $\kappa_{\rm e}$
are provided where appropriate. On the upper-left
panel (HD 209458b), models with the lighter gray shade(s) are for
the higher value(s) of P$_n$. Notice also that two different values
for the flux at 24 $\mu$m are shown on this same panel.
The one with the question mark is a tentative update to
the Deming et al. (2005) 24-$\mu$m measurement, kindly provided
by Drake Deming (private communication). If the flux at
24 $\mu$m is indeed $\sim$0.0033$\pm{0.0003}$, then our model(s)
with inversions provide the best fit at that wavelength as well.
(Taken from Burrows et al. 2008.)
}
\label{fig2}
\end{figure}

\begin{figure}
\begin{center}
 \includegraphics[width=2.4in,angle=270]{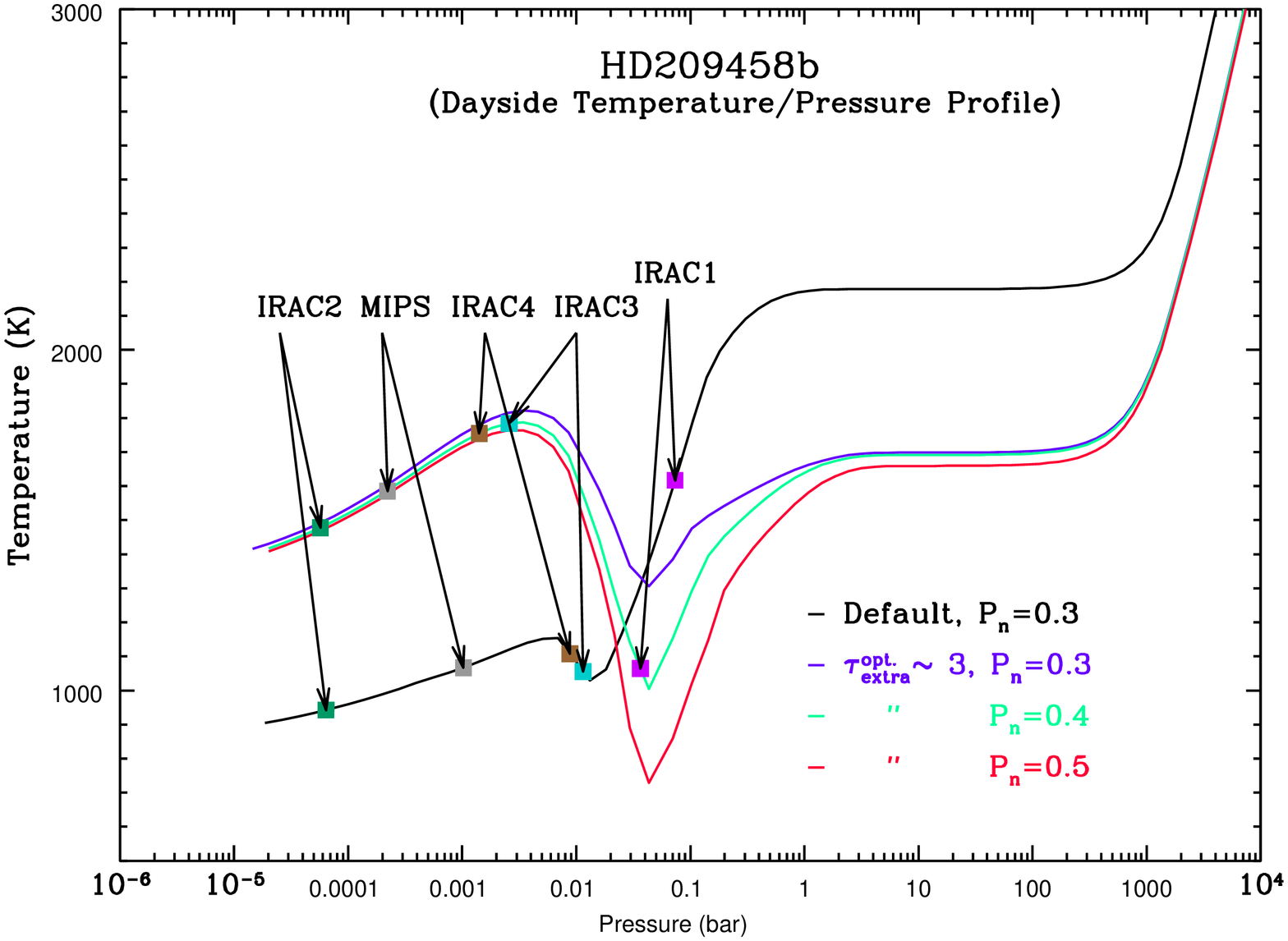} 
 \caption{Day-side temperature/pressure profiles for the four models for
HD209458b depicted in the upper left panel of
Fig. \ref{fig2}. We also indicate the position where the
monochromatic optical depths for the characteristic wavelengths of the four
IRAC and the MIPS channel are equal to 2/3.
(Taken from Burrows et al. 2007.)}
\label{fig3}
\end{center}
\end{figure}

The results are displayed in the four panels of Fig.~\ref{fig2}
for the four transiting planets. For HD 209458b, we show
three models with an upper-atmosphere extra absorber (and with resulting
stratospheres), and one traditional model without an extra absorber.
The values of the redistribution parameter
$P_n$ for the models with an extra absorber are 0.3, 0.4, and 0.5, and
the limiting pressures for the energy sink are 0.01 and 0.1 bars. The extra
absorber, assumed to contribute between 0.4 and 1 $\mu$m, is placed at altitude
below pressures of 25 mbars.
The corresponding $T/P$ profiles for the four HD 209458b models depicted
in Fig.~\ref{fig2}, together with the ``formation
depths," defined as the position in the atmosphere where the monochromatic
optical depth reaches 2/3, are displayed in Fig. \ref{fig3}. The plot explains why
the flux in the IRAC 2, 3, 4, and MIPS bands are higher for the model with an
extra absorber (stratosphere), while it is lower in the IRAC 1 band.

Figure \ref{fig2} shows that for HD209458b the models with 
a stratosphere fit the IRAC 1, 2, and 4
points quite well. However, the IRAC 3 point near 5.8 $\mu$m is difficult to
fit since the radiation in this band is formed at essentially the same position
in the atmosphere as the IRAC 4. Analogously, for HD149026b, the model with
a stratosphere is clearly preferable. The situation for HD 189733b
is uncertain; in any case a stratosphere may exist, but there is no 
compelling case for it. On the other hand, the TrES-1 data 
are fitted perfectly with a traditional model without a
stratosphere. Since the total irradiated flux is largest for
HD 149026b and smallest for TrES-1 -- see \cite{BBH08}, the presence of the 
stratosphere seems to be very roughly correlated with the total 
stellar flux at the planet's surface. 
In \cite{BBH08}, we have analyzed also the observed light curves of two
non-transiting planets ($\upsilon$ And b, and HD179949b), and found that for
$\upsilon$ And b the presence
of a stratosphere is strongly indicated in order to fit the observed data.

\section{Conclusions}

We have found (\cite{BHBKC08}; \cite{BBH08}) that a consistent fit to all data
at secondary eclipse for several strongly irradiated transiting planets
(HD 209458b, HD 149026b, and possibly HD 189733b), and very likely a
non-transiting planet $\upsilon$ And b, requires that their atmospheres
have temperature inversions -- stratospheres -- at altitude.
Such a thermal inversion affects:
(i) planet/star contrast ratios at the secondary eclipse;
(ii) their wavelength dependences; and
(iii) day-night flux contrast during a planetary orbit.
Moreover, the presence of the thermal inversion/stratosphere seems to 
roughly correlate with the total irradiated flux.

Temperature inversion is caused either by TiO/VO, as first suggested by
\cite{HBS03}, or by another, as yet unidentified,
opacity sources. These may be tholins, polyacetylenes, or various
non-equilibrium compounds. We invoke such extra absorbers, because
a cold-trap effect can operate to deplete the upper atmosphere of TiO/VO. 
However, one may speculate that with ongoing mass loss and/or 
rotational shear instabilities the atmosphere may be partially replenished 
in TiO/VO. Therefore, while TiO/VO might be responsible for the formation
of thermal inversions in the strongly irradiated planets, the exact nature
of the absorber must be viewed as very uncertain.


\begin{thebibliography}{}

\bibitem[Barman et al. (2001)]{BAH01} 
{Barman, T., Allard, F., \& Hauschildt, P.} (2001)
\textit{ApJ}, 556, 885

\bibitem[Burrows et al. (1997)]{Bur97}
{Burrows, A. et al.} 1997,
\textit{ApJ} 491, 875

\bibitem[Burrows \& Sharp (1999)]{BS99}
{Burrows, A., \& Sharp, C.} 1999,
\textit{ApJ} 512, 843

\bibitem[Burrows et al. (2004)]{BSH04}
{Burrows, A., Sudarsky, D., \& Hubeny, I.} 2004,
\textit{ApJ}, 609, 407

\bibitem[Burrows et al. (2006)]{BHS05}
{Burrows, A., Hubeny, I., \& Sudarsky, D.} 2005,
\textit{ApJ}, 625, L135

\bibitem[Burrows et al. (2006)]{BSH06}
{Burrows, A., Sudarsky, D., \& Hubeny, I.} 2006,
\textit{ApJ}, 650, 1140

\bibitem[Burrows et al. (2007)]{BHBKC08}
{Burrows, A., Hubeny, I, Budaj, J., Knutson, H. A., \& Charbonneau, D. } 2007,
\textit{ApJ}, 668, L171

\bibitem[Burrows et al. (2008)]{BBH08}
{Burrows, A., Budaj, J. \& Hubeny, I.} 2008a,
\textit{ApJ}, 678, 1436


\bibitem[Charbonneau et al. (2005)]{Ch05}
{Charbonneau, D. et al.} 2005,
\textit{ApJ}, 626, 523

\bibitem[Deming et al. (2005)]{De05}
{Deming, D., Seager, S., Richardson, L. J., \& Harrington, J.} 2005,
\textit{Nature}, 434, 740

\bibitem[Deming et al. (2006)]{De06}
{Deming, D., Harrington, L. J., Seager, S., Richardson, L. J.} 2006,
\textit{ApJ}, 644, 560

\bibitem[Fortney et al. (2006)]{For06}
{Fortney, J. J., Saumon, D., Marley, M. S., Lodders, K., \& Freedman, R. S.} 2006,
\textit{ApJ}, 642, 495

\bibitem[Fortney et al. (2008)]{For08}
{Fortney, J. J., Lodders, K., Marley, M. S., \& Freedman, R. S.} 2008,
\textit{ApJ}, 678, 1419

\bibitem[Harrington et al. (2007)]{Har07}
{Harrington, L. J. et al.} 2007,
\textit{Nature}, 447, 691

\bibitem[Hubeny (1988)]{Hube88}
{Hubeny, I.} 1988,
\textit{Computer Phys. Commun.}, 52, 103

\bibitem[Hubeny \& Lanz (1995)]{HL95}
{Hubeny, I. \& Lanz, T.} 1995
\textit{ApJ}, 439, 875

\bibitem[Hubeny et al. (2003)]{HBS03}
{Hubeny, I., Burrows, A., \& Sudarsky, D.} 2003,
\textit{ApJ}, 594, 1011

\bibitem[Knutson et al. (2007)]{Kn07a}
{Knutson, H. A., Charbonneau, D., Noyes, R. W., Brown, T. M., 
Gilliland, R. L.} 2007,
\textit{ApJ}, 655, 564

\bibitem[Knutson et al. (2008)]{Kn08}
{Knutson, H. A., Charbonneau, D., Allen, L. E., Burrows, A., Megeath, S. T.} 2008
\textit{ApJ}, 673, 526

\bibitem[Konacki et al. (2003)]{Kon03}
{Konacki, M., Torres, G., Jha, S., \& Sasselov, D.} 2003, 
\textit{Nature}, 421, 507

\bibitem[Richardson et al. (2003)]{Ri03}
{Richardson, L. J., Deming, D., \& Seager, S.} 2003,
\textit{ApJ}, 597, 581

\bibitem[Richardson et al. (2007)]{Ri07}
{Richardson, L. J., Deming, D., Horning, K., Seager, S., \& Harrington, J.} 2007,
\textit{Nature}, 445, 892


\bibitem[Seager \& Sasselov (1998)]{SeSa98}
{Seager, S., \& Sasselov, D.} 1998
\textit{ApJ}, 502, L157

\bibitem[Sharp \& Burrows (2007)]{SB07}
{Sharp, C. \& Burrows, A.} 2007
\textit{ApJS}, 168, 140

\bibitem[Sudarsky et al. (2003)]{SBH03}
{Sudarsky, Burrows, A., \& Hubeny, I.} 2003,
\textit{ApJ}, 588, 1121

\bibitem[Sudarsky et al. (2000)]{SBP00}
{Sudarsky, Burrows, A., \& Pinto, P.} 2000,
\textit{ApJ}, 538, 885



\end{thebibliography}
\end{document}